\UseRawInputEncoding 

\documentclass[aps,pra,twocolumn,showpacs,superscriptaddress]{revtex4-1}
\usepackage{graphicx}
\usepackage{dcolumn}
\usepackage{bm}
\usepackage{color}
\usepackage{times}
\usepackage{amssymb}
\usepackage{amsmath}
\usepackage{epsfig}
\usepackage{epstopdf}
\usepackage{dsfont}
\usepackage{subfigure}
\usepackage[colorlinks=true,citecolor=blue, linkcolor=blue,urlcolor=blue]{hyperref}
\usepackage{bookmark}
\usepackage{color}

\newenvironment{proof}{{\indent  \it Proof:\,}}{\hfill $\blacksquare$\par}
\definecolor{Green3}{rgb}{0.80,0.87,0.76}

\begin{document}

\title{Quantifying genuine tripartite entanglement by \textcolor[rgb]{0.00,0.00,0.00}{reshaping the state}}

\author{Dong-Dong Dong}
 \affiliation{School of Physics and Optoelectronics Engineering, Anhui University, Hefei
230601,  People's Republic of China}
\author{Li-Juan Li}
\affiliation{School of Physics and Optoelectronics Engineering, Anhui University, Hefei
  230601,  People's Republic of China}
\author{Xue-Ke Song}%
 \affiliation{School of Physics and Optoelectronics Engineering, Anhui University, Hefei
230601,  People's Republic of China}

 \author{Liu Ye}
  \affiliation{School of Physics and Optoelectronics Engineering, Anhui University, Hefei
230601,  People's Republic of China}
\author{Dong Wang}
 \email{dwang@ahu.edu.cn}
  \affiliation{School of Physics and Optoelectronics Engineering, Anhui University, Hefei
230601,  People's Republic of China}

\date{\today}

\begin{abstract}
Although genuine multipartite entanglement (GME), as one quantum resource, is indispensable in quantum information processing, most of the existing measures cannot detect GME faithfully.
In this \textcolor[rgb]{0.00,0.00,0.00}{paper},
 we present a novel GME measure, namely the minimum pairwise concurrence (MPC), by introducing  pairwise entanglement, which
    characters the entanglement between two single-qubit subsystems of a multipartite system without tracing out the remaining qubit.
   \textcolor[rgb]{0.00,0.00,0.00}{The pairwise entanglement can be obtained by combining the entanglement of reduced subsystem and three-tangle.}
      Compared with the   existing measures, the MPC measure  outperforms the previous ones in many aspects.
 Due to its fine properties, it thus is believed that the MPC could be one of  good candidates in achieving  potential quantum tasks and also facilitate the understanding for  GME.
\end{abstract}

\maketitle

\section{Introduction}
Entanglement, as one of the most fundamental features of quantum mechanics, describes the tensorial nonbiseparability of states for two or more parties. It is an indispensable resource in realistic quantum information processing, such as quantum cryptography \cite{PhysRevLett.67.661}, quantum teleportation \cite{PhysRevLett.70.1895}, and
dense coding \cite{PhysRevLett.69.2881}. Therefore, how to quantify entanglement has attracted considerable attention \cite{PhysRevLett.78.2275,10.1063/1.1494474}.
Entanglement measures for two-qubit states have been well studied, such as concurrence \cite{PhysRevLett.78.5022}, entanglement of formation \cite{doi:10.1080/09500349908231260} and negativity \cite{PhysRevA.65.032314}. In fact, all these  measures are equivalent in the sense that they all give the same entanglement ordering for different states \cite{Singh:20}.

Multipartite entanglement tends to be more effective in quantum information tasks.
 For example, three-party teleportation was shown to be  more robust to cheating and eavesdropping than two-party one \cite{PhysRevA.59.1829}. However, the quantification of  multipartite entanglement becomes much more complicated.
 It is not feasible to establish a consistent entanglement ordering using monotonicity, even for three-qubit systems \cite{PhysRevA.62.062314}. In the past few decades, although a series of multipartite entanglement measures have been proposed \cite {HBarnum_2001,PhysRevA.64.022306,10.1063/1.1497700,PhysRevLett.93.230501,PhysRevA.61.052306,PhysRevLett.106.190502}, most of them are not genuine multipartite entanglement (GME) measure. The definition of GME measure was identified by Ma \emph{et al.} \cite{PhysRevA.83.062325}.

 {\it Definition 1.\,}
 A GME measure should satisfy at least four conditions:

(a) {\it The measure
is zero for all product and biseparable states.}

(b) {\it The measure is positive for all nonbiseparable
states.}

(c) {\it The measure is invariant under local unitary transformations.}

(d) {\it The measure is nonincreasing under local operations and classical communication (LOCC).}

For example, the famous three-tangle \cite{PhysRevA.61.052306} violates condition (b), thus it is not a GME measure.
Recently, several attempts have been made to formulate GME measures \cite{PhysRevA.83.062325,PhysRevA.81.012308,PhysRevA.106.042415,PhysRevA.107.052403,PhysRevLett.127.040403,PhysRevResearch.4.023059}.
Explicitly, the genuinely multipartite concurrence (GMC) is defined as the minimum bipartite concurrence for multipartite pure states \cite{PhysRevA.83.062325}.
The generalized geometric measure (GGM) is based on the geometric distance
between the multipartite state and the set of all states that are not genuinely entangled \cite{PhysRevA.81.012308}, which is verified to be equivalent to GMC for any three-qubit pure states \cite{PhysRevA.106.042415}. For three-qubit states, concurrence fill was related to  the area of  concurrence triangle \cite{PhysRevLett.127.040403}. The  geometric mean of bipartite concurrences (GBC) was introduced as the geometric mean of regularized bipartite concurrences, which characters the entanglement of all possible bipartitions for a multipartite pure state \cite{PhysRevResearch.4.023059}.
However, all these measures cannot fully depict the nature of GME, e.g.,  the essence of the GMC is determined by bipartite entanglement. Concurrence fill does not satisfy condition (d) of GME measure \cite{PhysRevA.107.032405}. The values of Concurrence fill and GBC may greater than those of bipartite entanglement, which is counter-intuitive.

In this article, we present a new GME measure, namely the minimum pairwise concurrence (MPC), completely different from  the  previous GME measures being via bipartite entanglement. Particularly, our proposed GME measure is based on ``pairwise entanglement'' (PE), which is the entanglement between two single-qubit subsystems of a multipartite system without tracing out the remaining qubit.
\textcolor[rgb]{0.00,0.00,0.00}{The PE can be attained by combining the entanglement of reduced subsystem and three-tangle.}
In addition, we identify several existing GME measures and  demonstrate the superiority of our newly proposed measure over them.
\textcolor[rgb]{0.00,0.00,0.00}{Finally, the MPC is extended to the case of mixed states, and illustrate  two examples for clarity.}
The MPC holds the potential for extension to multipartite systems, and enhances   our comprehension for GME.

\section{Pairwise entanglement}
The key of quantifying genuine multipartite entanglement (GME) is how to identify a separable state. A pure $n$-partite state $\left| \Psi  \right\rangle $ is   biseparable if it can be written as
\begin{align}
\left| \Psi  \right\rangle  = \left| {{\phi _1}} \right\rangle  \otimes \left| {{\phi _2}} \right\rangle,
  \label{Eq.1}
    \end{align}
where $\left| {{\phi _1}} \right\rangle $ or $\left| {{\phi _2}} \right\rangle $ corresponds to a single subsystem or a group of subsystems. If the state $\left| \Psi  \right\rangle $ cannot be written as such form, then
 it is called as a genuinely $n$-partite entanglement. Specifically, a GME measure must satisfy Definition 1.
The identification of separable state can be simplified if one knows the entanglement between two single-qubit subsystems of a state, which is called as pairwise entanglement.
For a $n$-partite state, it can be divided into two partitions, one contains qubit $A$ while the other has $B$. One can conclude that the two partitions are entangled if  the PE
of $(A,B)$ exists.
Analogous to GME measure, the PE measure should  satisfy two additional conditions except for (c) and (d):

 ($a'$) {\it The PE measure of $(A,B)$ must be zero, once there exists  two divided partitions containing $A$ and $B$ respectively which are separable, as to a $n$-partite state.}

 ($b'$) {\it The measure is  positive if there are no such  states satisfying condition  ($a'$).}

 Unfortunately, the reduced density matrix fails to depict PE. For example,
     Greenberger-Horne-Zeilinger (GHZ) state $1/\sqrt 2 \left( {\left| {000} \right\rangle  + \left| {111} \right\rangle } \right)$ is the maximally three-qubit entangled state. However, if one of its
three qubits is traced out, the remaining state is completely disentangled.
That is to say, the  correlation between $A$ and $B$ is ignored by means of the reduced density matrix. Herein, we  naturally raise an interesting question:
how to character PE without loss of other subsystems?

To address the posed inquiry,
\textcolor[rgb]{0.00,0.00,0.00}{we consider the monogamy relation of concurrence for three-qubit system \cite{PhysRevA.61.052306} }
  \begin{align}
  C_{A B}^{2}+C_{A C}^{2} \leqslant C_{A(B C)}^{2},
 \label{Eq.15}
    \end{align}
    where $C_{A B}$ and $C_{A C}$  represent the entanglements  of reduced subsystems, both of which trace out one qubit.
    From this inequity, the three-tangle (or residual entanglement) is defined as \cite{PhysRevA.61.052306}
     \begin{align}
  {\tau _{ABC}} = C_{A\left( {BC} \right)}^2 - C_{AB}^2 - C_{AC}^2.
 \label{Eq.16}
 \end{align}

 \textcolor[rgb]{0.00,0.00,0.00}{It is known that ${C_{A(BC)}} $ represents the entanglement between $A$ and $BC$,  which includes the entanglements of reduced subsystems $\rho_{AB}$ and $\rho_{AC}$, and three-tangle.
 Note that, there  still exists entanglement between $A$ and $B$ in three-tangle. That is to say, PE should contain the entanglement of reduced subsystem and  three-tangle. }

 \textcolor[rgb]{0.00,0.00,0.00}{
 {\it Proposition 1.}
For an arbitrary three-qubit pure state, the PE between  $A$ and $B$ can be characterized by
 \begin{align}
  \mathcal{C}_{A^{'} B^{'}}=\sqrt {C_{AB}^2 + {\tau _{ABC}}},
 \label{Eq.4a}
    \end{align}
    which is called as pairwise concurrence.
Similarly, the pairwise concurrence between  $A$ and $C$ is given  by $\mathcal{C}_{A^{'} C^{'}}=\sqrt {C_{AC}^2 + {\tau _{ABC}}}$, and the pairwise concurrence between  $B$ and $C$ is given  by $\mathcal{C}_{B^{'} C^{'}}=\sqrt {C_{BC}^2 + {\tau _{ABC}}}$.
}

\begin{proof}
Without loss of generality, for product state of $A$ and $BC$,
\textcolor[rgb]{0.00,0.00,0.00}{ we have
\begin{align}
C_{A\left( {BC} \right)}=  C_{AB} = C_{AC}=0.
 \label{Eq.5a}
    \end{align}
Thus $\mathcal{C}_{A^{'} B^{'}}=0$,
}
which satisfies condition ($a'$).
For condition ($b'$),
consider the generalized Schmidt decomposition of three-qubit states \cite{PhysRevLett.85.1560}
    \begin{align}
|\psi\rangle_s=\lambda_{0}|000\rangle \!+\! \lambda_{1} e^{i \varphi}|100\rangle\!+\!\lambda_{2}|101\rangle\!+\!\lambda_{3}|110\rangle\!+\!\lambda_{4}|111\rangle,
 \label{Eq.4}
    \end{align}
    where $\lambda _{i}\geqslant 0$, $0 \leqslant \varphi  \leqslant \pi $ and $\sum\nolimits_i {\lambda _i^2 = 1} $.
     The bipartite entanglement  of $A$ and $BC$ is given by
  \begin{align}
{C_{A(BC)}} = 2{\lambda _0}\sqrt { \lambda _2^2+ \lambda _3^2+ \lambda _4^2} .
 \label{Eq.11}
    \end{align}
    Similarly,
     \begin{align}
{C_{B\left( {AC} \right)}} \!=\! 2\sqrt {\lambda _0^2\left( {\lambda _3^2 \!+\! \lambda _4^2} \right)\! +\! \lambda _1^2\lambda _4^2 \!+\! \lambda _2^2\lambda _3^2 \!-\! 2{\lambda _1}{\lambda _2}{\lambda _3}{\lambda _4}\cos \varphi }  .
 \label{Eq.12}
    \end{align}
  \textcolor[rgb]{0.00,0.00,0.00}{   The pairwise concurrence  of subsystem $AB$ is
        \begin{align}
\mathcal{C}_{A^{'} B^{'}} = 2{\lambda _0}\sqrt {\left( {\lambda _3^2 + \lambda _4^2} \right)} .
 \label{Eq.13}
    \end{align}}
    From ${C_{A(BC)}} > 0$ and ${C_{B(AC)}} > 0$, we have ${\lambda _0} > 0$ and   $\lambda _3$ and $\lambda _4$ cannot be 0 simultaneously, implying $\mathcal{C}_{A^{'} B^{'}} > 0$. As a result, the condition ($b'$) is satisfied.
    \textcolor[rgb]{0.00,0.00,0.00}{As three-tangle and the concurrence of any subsystem have been proven to be invariant under local unitary transformations and nonincreasing under LOCC \cite{PhysRevA.62.062314,PhysRevLett.95.260502},   $\mathcal{C}_{A^{'} B^{'}}$ certainly satisfies the conditions (c) and (d).
    }
\end{proof}

\begin{figure}
    \centering
    \includegraphics[width=8.6cm]{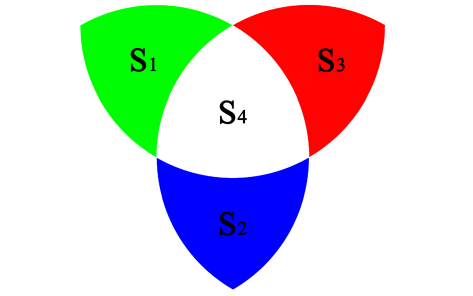}
    \caption{(Color online)  The relations among pairwise concurrences and bipartite concurrences.
    ${S_1} + {S_4}$ represents the pairwise concurrence squared  $\mathcal{C}_{A^{'} B^{'}}^2$. ${S_2} + {S_4}$ represents the pairwise concurrence  squared $\mathcal{C}_{A^{'} C^{'}}^2$. ${S_4}$ is the shareable tripartite entanglement (three-tangle), and $S_1+{S_2} + {S_4}$ represents the bipartite concurrence squared ${C_{A(BC)}^2}$.
    }
    \label{f1}
  \end{figure}

\textcolor[rgb]{0.00,0.00,0.00}{
From Proposition 1, one can easily obtain a relation among pairwise concurrences and bipartite concurrence
   \begin{align}
  \mathcal{C}_{A^{'} B^{'}}^{2}+\mathcal{C}_{A^{'} C^{'}}^{2} - C_{A(B C)}^{2}={\tau _{ABC}},
 \label{Eq.17}
    \end{align}
    whose form is opposite to that in Eq. (\ref{Eq.16}).
In fact, such a result is reasonable.
Bipartite concurrence $C_{A(B C)}$ includes pairwise concurrences $\mathcal{C}_{A^{'} B^{'}}$ and $\mathcal{C}_{A^{'} C^{'}}$, which contain three-tangle. Specifically, three-tangle, as shareable tripartite entanglement in pairwise concurrences, is  added twice in $\mathcal{C}_{A^{'} B^{'}}^{2}+\mathcal{C}_{A^{'} C^{'}}^{2}$,
} as shown in Fig. \ref{f1}.

\textcolor[rgb]{0.00,0.00,0.00}{
In addition, we find a simpler method to calculate pairwise concurrence  for the  generalized Schmidt decomposition of three-qubit states.}
Firstly, from the state we know the probability distribution of different measure outcomes of subsystem AB.
Then, we construct a state by taking the square root of that probability distribution, which is called reshaped state.
\textcolor[rgb]{0.00,0.00,0.00}{The concurrence  of the reshaped state is equal  to $\mathcal{C}_{A^{'} B^{'}}$.}
Typically, the concurrence  can be expressed as a simple formula  for two-qubit pure states
     \begin{align}
  C\left({\left| \phi  \right\rangle } \right) = 2\left| {ps - qr} \right|,
 \label{Eq.3}
    \end{align}
where  ${\left| \phi  \right\rangle } = p\left| {00} \right\rangle  + q\left| {01} \right\rangle  + r\left| {10} \right\rangle  + s\left| {11} \right\rangle $.

\textcolor[rgb]{0.00,0.00,0.00}{{\it Proposition 2.}}
For the generalized Schmidt decomposition of three-qubit states
    \begin{align}
|\psi\rangle_s=\lambda_{0}|000\rangle \!+\! \lambda_{1} e^{i \varphi}|100\rangle\!+\!\lambda_{2}|101\rangle\!+\!\lambda_{3}|110\rangle\!+\!\lambda_{4}|111\rangle, \nonumber
    \end{align}
\textcolor[rgb]{0.00,0.00,0.00}{the pairwise concurrence  $\mathcal{C}_{A^{'} B^{'}}$ is equal to ${\cal C}(\left| \psi  \right\rangle _{AB})$}, with
  \begin{align}
 {\left| \psi  \right\rangle _{AB}} = {\lambda _0}\left| {00} \right\rangle  + \sqrt {\lambda _1^2 + \lambda _2^2} \left| {10} \right\rangle  + \sqrt {\lambda _3^2 + \lambda _4^2} \left| {11} \right\rangle.
 \label{Eq.5}
    \end{align}
\textcolor[rgb]{0.00,0.00,0.00}{The pairwise concurrence  $\mathcal{C}_{A^{'} C^{'}}$ is equal to ${\cal C}(\left| \psi  \right\rangle _{AC})$}, with
     \begin{align}
 {\left| \psi  \right\rangle _{AC}} = {\lambda _0}\left| {00} \right\rangle  + \sqrt {\lambda _1^2 + \lambda _3^2} \left| {10} \right\rangle  + \sqrt {\lambda _2^2 + \lambda _4^2} \left| {11} \right\rangle.
 \label{Eq.6}
    \end{align}
However, \textcolor[rgb]{0.00,0.00,0.00}{the pairwise concurrence  $\mathcal{C}_{B^{'} C^{'}}$ is equal to ${\cal C}(\left| \psi  \right\rangle _{BC})$}, where
     \begin{align}
 {\left| \psi  \right\rangle _{BC}} \!=\! \left| {{l_0}} \right|\left| {00} \right\rangle \! +\! \sqrt {{{\left| {{l_1}} \right|}^2} \!+\! {{\left| {{l_3}} \right|}^2}} \left| {10} \right\rangle  \!+\! \sqrt {{{\left| {{l_2}} \right|}^2} \!+ \! {{\left| {{l_4}} \right|}^2}} \left| {11} \right\rangle .
 \label{Eq.7}
    \end{align}
    State $\left| \psi  \right\rangle _{BC}$ is the reshaped state of $|\psi\rangle_1$
    \begin{align}
|\psi\rangle_1=l_{0}|000\rangle \!+\! l_{1} |010\rangle\!+\!l_{2}|011\rangle\!+\!l_{3}|110\rangle\!+\!l_{4}|111\rangle,
 \label{Eq.8}
    \end{align}
    which is transformed from $|\psi\rangle_s$ by  a group of  local unitary transformations
    \begin{widetext}
\begin{align}
\begin{split}
  {U_A} = \left( {\begin{array}{*{20}{c}}
0&1\\
1&0
\end{array}} \right),\quad {U_B} = \left( {\begin{array}{*{20}{c}}
0&{ - 1}\\
1&0
\end{array}} \right), \quad
{U_C} = \frac{{ - 1}}{{\sqrt {\lambda _3^2 + \lambda _4^2} }}\left( {\begin{array}{*{20}{c}}
{{\lambda _3}}&{{\lambda _4}}\\
{{\lambda _4}}&{ - {\lambda _3}}
\end{array}} \right),
\end{split}
 \label{Eq.23}
    \end{align}
    i.e.,
    \begin{align}
  |\psi {\rangle _1} = \sqrt {\lambda _3^2 + \lambda _4^2} |000\rangle  - \frac{{{\lambda _1}{\lambda _3}{e^{i\varphi }} + {\lambda _2}{\lambda _4}}}{{\sqrt {\lambda _3^2 + \lambda _4^2} }}|010\rangle  + \frac{{{\lambda _2}{\lambda _3} - {\lambda _1}{\lambda _4}{e^{i\varphi }}}}{{\sqrt {\lambda _3^2 + \lambda _4^2} }}|011\rangle  - \frac{{{\lambda _0}{\lambda _3}}}{{\sqrt {\lambda _3^2 + \lambda _4^2} }}|110\rangle  - \frac{{{\lambda _0}{\lambda _4}}}{{\sqrt {\lambda _3^2 + \lambda _4^2} }}|111\rangle .
 \label{Eq.24}
    \end{align}
    \end{widetext}

\textcolor[rgb]{0.00,0.00,0.00}{This proposition can be proved  by direct calculation. }

\section{The minimum pairwise concurrence (MPC)}
Using three pairwise concurrences, we now present a  genuine tripartite entanglement measure.

{\it Theorem 1.\,}  For  an arbitrary three-qubit pure state, the MPC is a genuine tripartite entanglement measure, which is formulated  by
\textcolor[rgb]{0.00,0.00,0.00}{\begin{align}
{{\cal M}}_{ABC} = \min \left\{ \mathcal{C}_{A^{'} B^{'}},\mathcal{C}_{A^{'} C^{'}},\mathcal{C}_{B^{'} C^{'}} \right\}.
 \label{Eq.14}
    \end{align}}

    The proof is straightforward. Since the pairwise concurrences satisfy conditions ($a'$), ($b'$), (c) and (d), then the MPC satisfies Definition 1.
  In addition, Xie \emph{ et al.} proposed a new condition that a  genuine tripartite entanglement measure should satisfy \cite{PhysRevLett.127.040403}: (e) the measure ranks the GHZ state
as more entangled than  W state $1/\sqrt 3 \left( {\left| {001} \right\rangle  + \left| {010} \right\rangle  + \left| {100} \right\rangle } \right)$.
The reason is that,  in three-party
teleportation,  the GHZ state  can faithfully teleport an
arbitrary single-qubit quantum state, while
the W state can only achieve a maximal success rate of 2/3 \cite{Joo_2003}.
It is a fact that the MPC conforms to condition  (e) as well.
Explicitly, the  MPC is maximized for the GHZ state, i.e., ${\cal M}\left( {\left| {GHZ} \right\rangle } \right) = 1$. Interestingly,  the  MPC happens to be 2/3 for the W state.

To further highlight the effectiveness of MPC, we can take the analogy in Ref. \cite{PhysRevA.67.052107}.
 For an ordinary chain with $N$ links, if $N -1$ of these are strong and the rest  one is weak, then the chain is close to breaking and so only has a small amount of ``nonbrokenness''. Analogously, a  $N$-partite pure state,
the  $N -1$ members of which  are strongly entangled and weakly correlated with the rest one,  is very close to possessing no genuinely $N$-partite entanglement.
 In this sense, it claims that a GME measure  should satisfy  a new condition:
 (f) the measure should guarantee that  GME is less than or equal to any bipartite entanglement.
 Without loss of generality,
 from Eqs. (\ref{Eq.11}), (\ref{Eq.12}) and (\ref{Eq.13}), one can see that $\mathcal{C}_{A^{'} B^{'}} \leqslant {C_{A\left( {BC} \right)}}$ and $\mathcal{C}_{A^{'} B^{'}} \leqslant {C_{B\left( {AC} \right)}}$, which indicates that the MPC satisfies condition (f) in the three-qubit scenarios.

   On the other hand,  most of GME measures are based on bipartite concurrence, including GMC, concurrence fill and GBC.
GMC is the minimum bipartite concurrence, concurrence fill is the area of bipartite concurrence triangle, and GBC is the geometric mean of bipartite concurrences. All of them satisfy some good properties such as discriminance, convexity and condition (e), where discriminance denotes the conditions (a) and (b). However, concurrence fill does not satisfy  entanglement monotone, and  may increase under LOCC on average \cite{PhysRevA.107.032405}. Concurrence fill and GBC do not satisfy condition (f), i.e., both of them cannot guarantee that the GME is less than or equal to bipartite entanglement. Note that, neither triangle measure nor geometric mean have clear physical significance. Besides, the partial   properties of  these GME measures have been provided in   Table \ref
{tab:table1}.

Basically, MPC and GMC are two inequivalent measures for tripartite states, since they may give rise to different entanglement orderings for  one specific pair of tripartite states. As an example, consider the following two states:
\begin{align}
\begin{split}
    &{\left| \psi  \right\rangle _5} = \frac{1}{{\sqrt {10} }}\left| {000} \right\rangle  + \frac{2}{{\sqrt 5 }}\left| {101} \right\rangle  + \frac{1}{{\sqrt {10} }}\left| {110} \right\rangle, \\
  &  {\left| \psi  \right\rangle _6} = \frac{3}{{\sqrt {20} }}\left| {000} \right\rangle  + \frac{1}{{\sqrt {10} }}\left| {101} \right\rangle  + \frac{3}{{\sqrt {20} }}\left| {110} \right\rangle .
 \label{Eq.30}
 \end{split}
    \end{align}
the entanglement for the above states is identical to  0.6 for using GMC measure. However, the entanglement for ${{{\left| \psi  \right\rangle }_5}}$ as 0.2 and the entanglement  for ${{{\left| \psi  \right\rangle }_6}}$ is 0.424, for MPC measure. Thus, MPC can distinguish the entanglement of different states that GMC cannot, manifesting the performance of MPC outperforms that of GMC to some extent.

\begin{table}[]
\caption{\label{tab:table1}%
Partial properties of various GME measures. They all can distinguish  entangled states from  separated ones. Concurrence fill is not an entanglement monotone. Concurrence and GBC can not guarantee  GME less than or equal to bipartite entanglement.
 }
\begin{ruledtabular}
\begin{tabular}{ccddd}
&Discriminance&
\multicolumn{1}{c}{\textrm{Monotonicity}}&
\multicolumn{1}{c}{\textrm{Condition (f)}}\\
\hline
GMC&\checkmark&\mbox{\checkmark}&\mbox{\checkmark}\\
$F_{123}$\footnotemark[1] & \checkmark & \times  & \times \\
GBC& \checkmark & \checkmark  & \times \\
MPC
  & \checkmark
  & \checkmark & \checkmark \\
\end{tabular}
\end{ruledtabular}
\footnotetext[1]{Concurrence fill}
\end{table}

\textcolor[rgb]{0.00,0.00,0.00}{\section{Generalization to mixed states}
Three-tangle of mixed three-qubit states has been defined by convex roof construction \cite{PhysRevA.77.032310}
\begin{align}
    {\cal \tau}(\rho)=\min _{\left\{p_{i}, \psi_{i}\right\}} \sum_{i} p_{i} {\cal \tau}\left(\psi_{i}\right),
 \label{Eq.29}
    \end{align}
    where the minimum is taken over all possible decompositions $\rho=\sum_{i } p_{i}\left|\psi_{i}\right\rangle\left\langle\psi_{i}\right|$.
    On the other hand, the entanglement of reduced subsystem itself satisfies the property of convexity \cite{PhysRevLett.80.2245}.
Therefore, the MPC can be generalized to the case of mixed states as
    \begin{align}
    {\cal M}(\rho)=\sqrt {{{\left( {\min \left[ {{C_{AB}},{C_{AC}},{C_{BC}}} \right]} \right)}^2} + \tau \left( \rho  \right)} .
 \label{Eq.29}
    \end{align}
In the following, we will focus on the MPC for rank-2 and rank-3 mixed states as illustrative examples to show the performance of our finding.
\subsection{Mixture of GHZ and W states}
 Lohmayer \emph{et al.} has shown how to construct the optimal decomposition for three-tangle for the rank-2 mixture of GHZ and W states \cite{PhysRevLett.97.260502}:
 \begin{align}
    \rho(p)=p|\mathrm{GHZ}\rangle\langle\mathrm{GHZ}|+(1-p)| {\rm W}\rangle\langle {\rm W}|,
 \label{Eq.22a}
    \end{align}
with the state's parameter $p\in[0,1]$.    The resulting  three-tangle for $\rho(p)$  has three different expressions depending on the range of $p$ as follows:
    \begin{align}
    \tau(\rho(p))=\left\{\begin{array}{ll}
0 & \text { for } 0 \leqslant p \leqslant p_{1}, \\
g_{I}(p) & \text { for } p_{1} \leqslant p \leqslant p_{2} ,\\
g_{I I}(p) & \text { for } p_{2} \leqslant p \leqslant 1,
\end{array}\right.
 \label{Eq.23a}
    \end{align}
    where
     \begin{align}
    g_{I}(p)&=p^{2}-\frac{8 \sqrt{6}}{9} \sqrt{p(1-p)^{3}}, \nonumber\\
        g_{I I}(p)&=1-(1-p)\left(\frac{3}{2}+\frac{1}{18} \sqrt{465}\right),\nonumber \\
      p_{1}&=\frac{4 \sqrt[3]{2}}{3+4 \sqrt[3]{2}} \sim 0.6269, \nonumber \\
       p_{2}&=\frac{1}{2}+\frac{3}{310} \sqrt{465} \sim 0.7087.
 \label{Eq.24a}
    \end{align}
    On the other hand, $C_{AB}$, $C_{AC}$ and $C_{BC}$ for state $\rho(p)$ are the same. $C_{AB}$ is given by
     \begin{align}
   C_{AB} =\left\{\begin{array}{ll}
\frac{2}{3}(1 - p) - \sqrt {\frac{p}{3}(2 + p)}  & \text { for } 0 \leqslant p \leqslant p_{0} ,\\
0 & \text { for } p_{0} \leqslant p \leqslant 1,
\end{array}\right.
 \label{Eq.25a}
    \end{align}
    where $p_{0}=7 - 3\sqrt 5 \sim 0.2918 $. As a result, we obtain the MPC  of  state $\rho(p)$
    \begin{align}
{\cal M}(\rho(p))=\left\{\begin{array}{ll}
\frac{2}{3}(1 - p) - \sqrt {\frac{p}{3}(2 + p)} & \text { for } 0 \leqslant p \leqslant p_{0}, \\
0 & \text { for } p_{0} \leqslant p \leqslant p_{1}, \\
\sqrt {g_{I}(p)} & \text { for } p_{1} \leqslant p \leqslant p_{2} ,\\
\sqrt {g_{I I}(p)} & \text { for } p_{2} \leqslant p \leqslant 1.
\end{array}\right.
 \label{Eq.26a}
    \end{align}
}

\begin{figure}
    \centering
    \includegraphics[width=8cm]{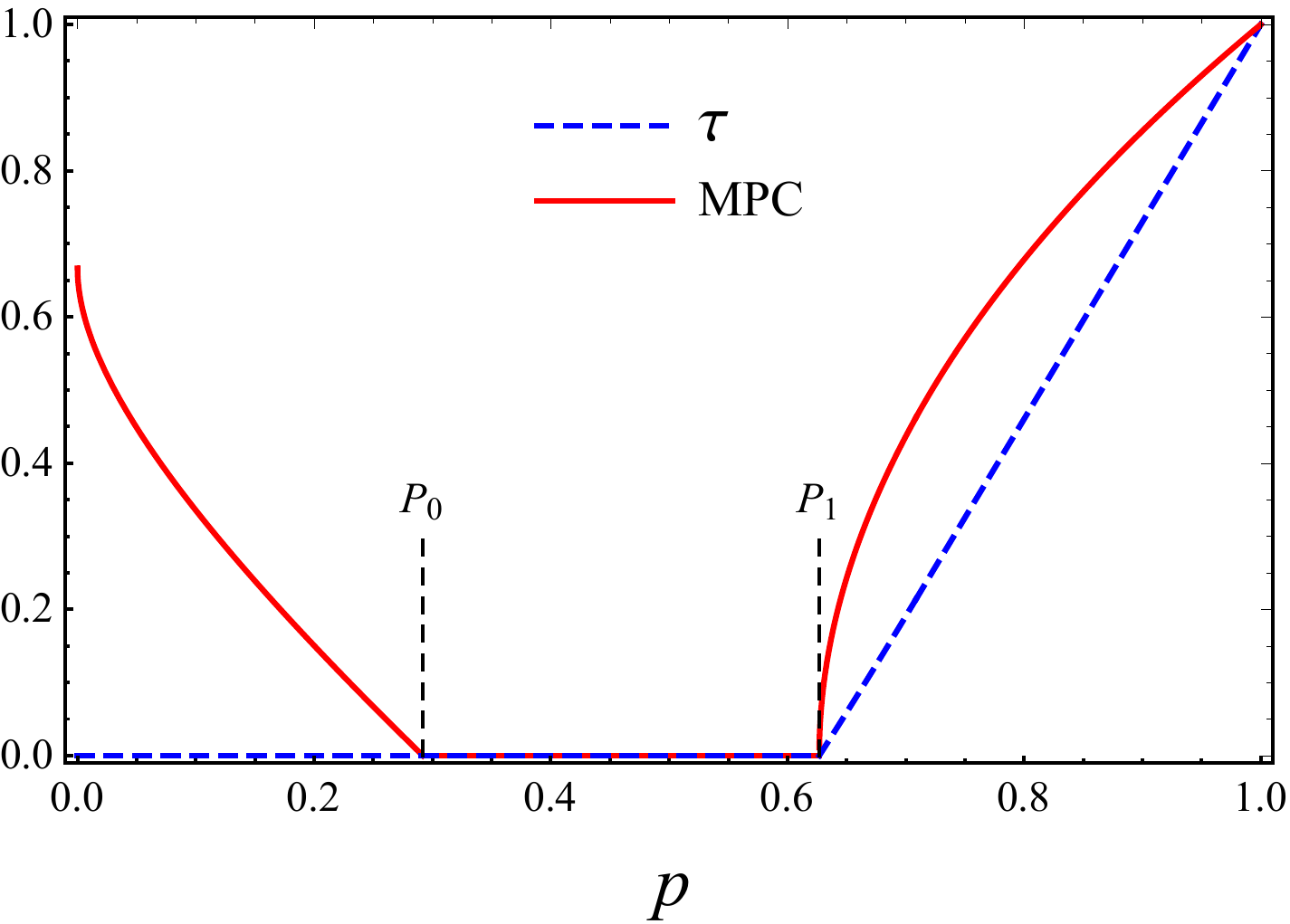}
    \caption{(Color online)  \textcolor[rgb]{0.00,0.00,0.00}{Plot of the MPC (solid line) and three-tangle (dashed line) for the state $\rho(p)$ as a function of the parameter $p$.
    }}
    \label{f2}
  \end{figure}

\textcolor[rgb]{0.00,0.00,0.00}{
Fig. \ref{f2} plots  the MPC and three-tangle change as a function of the parameter $p$ for $\rho(p)$. The results show that the MPC is always greater than or equal to three-tangle, and it can detect entanglement that the three-tangle cannot in the interval $0\leqslant p\leqslant p_0$. Thus, we say the proposed MPC outperforms the previous one when measuring the multipartite entanglements in the case. In addition, there is  no GME in the interval $p_0\leqslant p \leqslant p_1$.
}

\textcolor[rgb]{0.00,0.00,0.00}{
\subsection{Mixture of GHZ, W  and  flipped-W states}
In Ref. \cite{PhysRevA.79.024306}, Jung \emph{et al.} investigated  the optimal decompositions for the rank-3 mixed states, i.e., the mixture of GHZ, W, and flipped-W states
\begin{align}
    \rho(p, q)\!= \!p|\mathrm{GHZ}\rangle\langle\mathrm{GHZ}| \!+ \! q| {\rm W}\rangle\langle {\rm W}| \!+\! (1-p-q)| \tilde{{\rm W}}\rangle\langle\tilde{{\rm W}}|,
     \label{Eq.27a}
    \end{align}
    where
   \begin{align}
    |\tilde{{\rm W}}\rangle=\frac{1}{\sqrt{3}}(|110\rangle+|101\rangle+|011\rangle),
     \label{Eq.28a}
    \end{align}
    and $q=(1-p)/n$ with a positive integer $n$. The three-tangle for $\rho(p, q)$ was  formulated as
    \begin{align}
    \tau(\rho(p, q))=\left\{\begin{array}{ll}
0 & \text { for } 0 \leqslant p \leqslant p_{1}, \\
\alpha_{I}(p) & \text { for } p_{1} \leqslant p \leqslant p_{2}, \\
\alpha_{I I}(p) & \text { for } p_{2} \leqslant p \leqslant 1,
\end{array}\right.
     \label{Eq.29a}
    \end{align}
    where
    \begin{align}
    \alpha_{I}(p)= & p^{2}-\frac{4 \sqrt{n-1}}{n} p(1-p)-\frac{4(n-1)}{3 n^{2}}(1-p)^{2} \nonumber  \\
& -\frac{8 \sqrt{6 n}\left[1+(n-1)^{3 / 2}\right]}{9 n^{2}} \sqrt{p(1-p)^{3}} , \nonumber
    \end{align}
    \begin{align}
    {\alpha _{II}}\left( p \right) = \frac{{p - {p_2}}}{{1 - {p_2}}} + \frac{{1 - p}}{{1 - {p_2}}}{\alpha _I}\left( {{p_2}} \right). \nonumber
    \end{align}
    For  simplicity, let's consider \( n \) to be 2.
 Then the parameters $p_1$ and $p_2$ are given by
    \begin{align}
    {p_1} = 0.75, \quad {p_2} = \frac{{2 + \sqrt 3 }}{4} \sim 0.9330.
    \end{align}
In addition, $C_{AB}$, $C_{AC}$ and $C_{BC}$ for the state $\rho(p,q)$ have the identical values. $C_{AB}$ is given by
   \begin{align}
   C_{AB} =\left\{\begin{array}{ll}
\xi \left( p \right)  & \text { for } 0 \leqslant p \leqslant p_{0} ,\\
0 & \text { for } p_{0} \leqslant p \leqslant 1,
\end{array}\right.
 \label{Eq.31a}
    \end{align}
    where
    \begin{align}
   \xi \left( p \right) \!=\! \frac{2}{3}(1 \!-\! p) \!-\! \frac{1}{{3n}}\sqrt {\left( {2 \!+\! p\left( {3n \!-\! 2} \right)} \right)\left( {2\left( {p \!-\! 1} \right) \!+\! n\left( {p \!+\! 2} \right)} \right)} ,
 \label{Eq.32a}
    \end{align}
    and ${p_0} = 0.25$. Therefore, the MPC  of  state $\rho(p,q)$ is obtained as
    \begin{align}
{\cal M}(\rho(p,q))=\left\{\begin{array}{ll}
\xi \left( p \right) & \text { for } 0 \leqslant p \leqslant p_{0}, \\
0 & \text { for } p_{0} \leqslant p \leqslant p_{1}, \\
\sqrt {\alpha_{I}(p)} & \text { for } p_{1} \leqslant p \leqslant p_{2} ,\\
\sqrt {\alpha_{I I}(p)} & \text { for } p_{2} \leqslant p \leqslant 1.
\end{array}\right.
 \label{Eq.33a}
    \end{align}
}

\begin{figure}
    \centering
    \includegraphics[width=8cm]{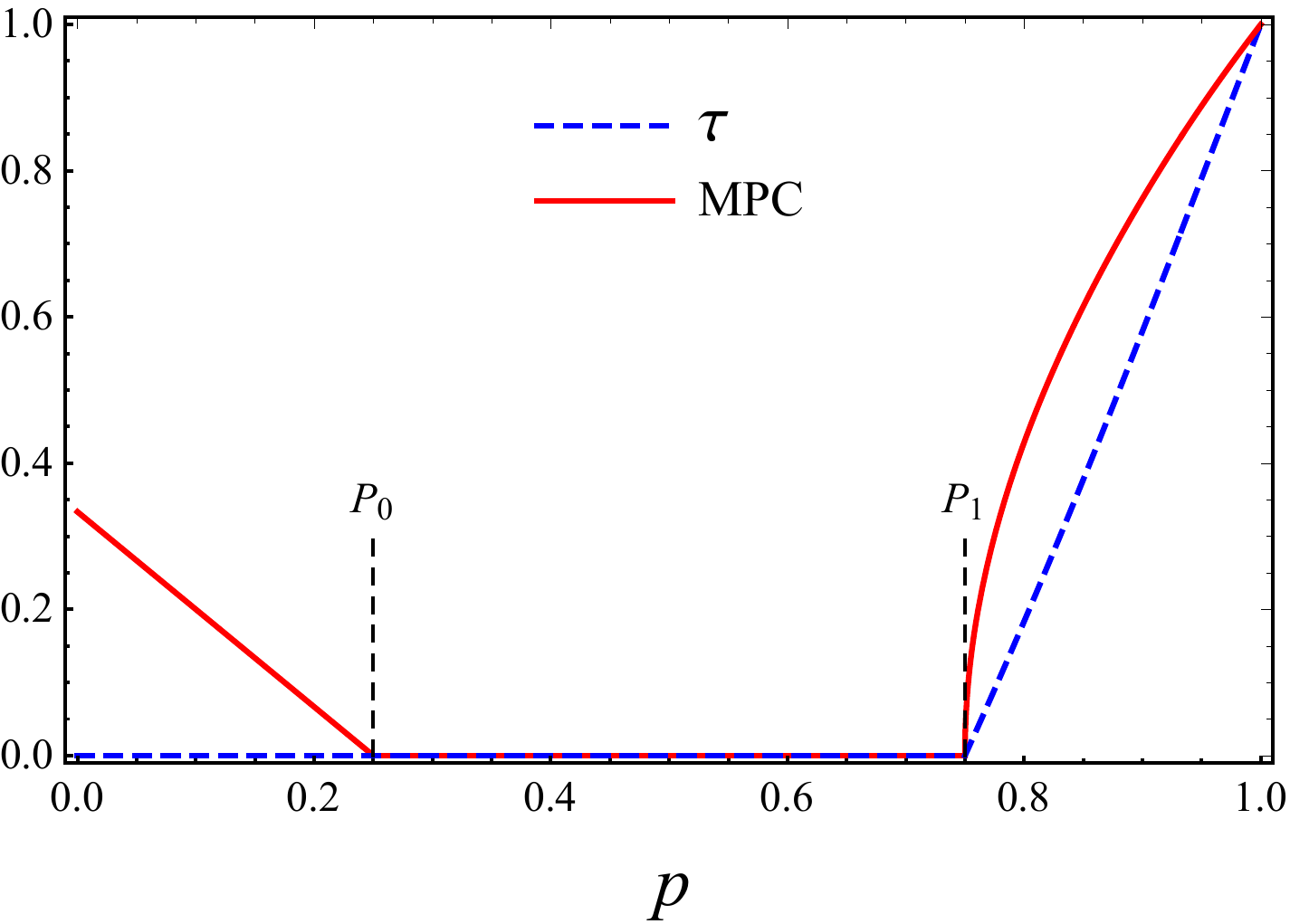}
    \caption{(Color online)  \textcolor[rgb]{0.00,0.00,0.00}{Plot of the MPC (solid line) and three-tangle (dashed line) in $\rho(p,q)$ as a function of $p$. The parameter q is taken to be $(1-p)/n$ and n is taken to be 2.
    }}
    \label{f3}
  \end{figure}

\textcolor[rgb]{0.00,0.00,0.00}{
Figure \ref{f3} plots how the MPC and three-tangle change with respect to parameter $p$ for $\rho(p,q)$. Following the figure,
it directly shows the MPC is always greater than or equal to the amount of three-tangle, and the MPC is capable of detecting entanglement within the range $0\leqslant p\leqslant p_0$ while the three-tangle is zero. This
implies that our measure is more powerful than the three-tangle for the entanglement detection in the current scenario, which virtually is in agreement with the conclusion made before.}

\section{Discussion and Summary}
\textcolor[rgb]{0.00,0.00,0.00}{The concept of PE} has  the potential to be generalized to $n$-qudit case. \textcolor[rgb]{0.00,0.00,0.00}{One feasible approach to quantify  it  is by utilizing the reshaped state as outlined in Proposition 2, where the entanglement of the reshaped state} should take the supremum over all local unitary transformations on the original multipartite state.
 Then once one obtains  \textcolor[rgb]{0.00,0.00,0.00}{the optimal reshaped states},
  the calculation times of PE are $P_n^2$, which is much less than $\sum\limits_{m = 1}^{[n/2]} {P_n^m} $ by virtue of  bipartite concurrences, where $P_{n}^{m}=\frac{n!}{m!(n-m)!}$ is the permutation. This will be  studied in the future,
   and there is a challenge of proving the monotonicity of the PE measure in the multipartite architectures.
    Remarkably,  the reshaping-state method used in Proposition 2 is available  for quantifying entanglement, thus it can be conjectured that the current reshaping-state methodology is effective to quantify  the other quantum resources in multipartite system, such  as Bell nonlocality and quantum steering.

In summary, we have proposed a new and effective approach of quantifying entanglement named as PE.
 The core idea of PE is to validly depict the entanglement between two single-qubit subsystems of a multipartite system without tracing out the remaining qubit.
  It verifies that the proposed MPC is a good measure of GME, which guarantees GME less than or equal to bipartite entanglement. Further, we compared the MPC with three other GME measures, including GMC, concurrence fill and GBC. The results show that,
the MPC outperforms the previous ones, which satisfies more conditions of GME measure than concurrence fill and GBC. Besides, it has been illustrated that the MPC is  superior to  GMC  in measuring certain types of entanglement.
\textcolor[rgb]{0.00,0.00,0.00}{Finally, the MPC has been generalized to mixed states, with two illustrative examples provided for clarity.}
   With these in mind, we argue  that  the proposed MPC  paves a new avenue for quantifying multipartite entanglement, which might help us better understand genuine multipartite entanglement in realistic quantum information processing.

\begin{acknowledgements}
We thank Dr. Z.-P. Xu  for the useful discussions.
  This work was supported by the National Science Foundation of China (Grant Nos. 12075001, 61601002, and 12004006), Anhui Provincial Key Research and Development Plan (Grant No. 2022b13020004), and Anhui Provincial Natural Science Foundation (Grant No. 2008085QA43).
\end{acknowledgements}


\begin{thebibliography}{0}%
\makeatletter
\providecommand \@ifxundefined [1]{%
 \@ifx{#1\undefined}
}%
\providecommand \@ifnum [1]{%
 \ifnum #1\expandafter \@firstoftwo
 \else \expandafter \@secondoftwo
 \fi
}%
\providecommand \@ifx [1]{%
 \ifx #1\expandafter \@firstoftwo
 \else \expandafter \@secondoftwo
 \fi
}%
\providecommand \natexlab [1]{#1}%
\providecommand \enquote  [1]{``#1''}%
\providecommand \bibnamefont  [1]{#1}%
\providecommand \bibfnamefont [1]{#1}%
\providecommand \citenamefont [1]{#1}%
\providecommand \href@noop [0]{\@secondoftwo}%
\providecommand \href [0]{\begingroup \@sanitize@url \@href}%
\providecommand \@href[1]{\@@startlink{#1}\@@href}%
\providecommand \@@href[1]{\endgroup#1\@@endlink}%
\providecommand \@sanitize@url [0]{\catcode `\\12\catcode `\$12\catcode
  `\&12\catcode `\#12\catcode `\^12\catcode `\_12\catcode `\%12\relax}%
\providecommand \@@startlink[1]{}%
\providecommand \@@endlink[0]{}%
\providecommand \url  [0]{\begingroup\@sanitize@url \@url }%
\providecommand \@url [1]{\endgroup\@href {#1}{\urlprefix }}%
\providecommand \urlprefix  [0]{URL }%
\providecommand \Eprint [0]{\href }%
\providecommand \doibase [0]{http://dx.doi.org/}%
\providecommand \selectlanguage [0]{\@gobble}%
\providecommand \bibinfo  [0]{\@secondoftwo}%
\providecommand \bibfield  [0]{\@secondoftwo}%
\providecommand \translation [1]{[#1]}%
\providecommand \BibitemOpen [0]{}%
\providecommand \bibitemStop [0]{}%
\providecommand \bibitemNoStop [0]{.\EOS\space}%
\providecommand \EOS [0]{\spacefactor3000\relax}%
\providecommand \BibitemShut  [1]{\csname bibitem#1\endcsname}%
\let\auto@bib@innerbib\@empty
\end{thebibliography}%


\begin{thebibliography}{32}%
\makeatletter
\providecommand \@ifxundefined [1]{%
 \@ifx{#1\undefined}
}%
\providecommand \@ifnum [1]{%
 \ifnum #1\expandafter \@firstoftwo
 \else \expandafter \@secondoftwo
 \fi
}%
\providecommand \@ifx [1]{%
 \ifx #1\expandafter \@firstoftwo
 \else \expandafter \@secondoftwo
 \fi
}%
\providecommand \natexlab [1]{#1}%
\providecommand \enquote  [1]{``#1''}%
\providecommand \bibnamefont  [1]{#1}%
\providecommand \bibfnamefont [1]{#1}%
\providecommand \citenamefont [1]{#1}%
\providecommand \href@noop [0]{\@secondoftwo}%
\providecommand \href [0]{\begingroup \@sanitize@url \@href}%
\providecommand \@href[1]{\@@startlink{#1}\@@href}%
\providecommand \@@href[1]{\endgroup#1\@@endlink}%
\providecommand \@sanitize@url [0]{\catcode `\\12\catcode `\$12\catcode
  `\&12\catcode `\#12\catcode `\^12\catcode `\_12\catcode `\%12\relax}%
\providecommand \@@startlink[1]{}%
\providecommand \@@endlink[0]{}%
\providecommand \url  [0]{\begingroup\@sanitize@url \@url }%
\providecommand \@url [1]{\endgroup\@href {#1}{\urlprefix }}%
\providecommand \urlprefix  [0]{URL }%
\providecommand \Eprint [0]{\href }%
\providecommand \doibase [0]{http://dx.doi.org/}%
\providecommand \selectlanguage [0]{\@gobble}%
\providecommand \bibinfo  [0]{\@secondoftwo}%
\providecommand \bibfield  [0]{\@secondoftwo}%
\providecommand \translation [1]{[#1]}%
\providecommand \BibitemOpen [0]{}%
\providecommand \bibitemStop [0]{}%
\providecommand \bibitemNoStop [0]{.\EOS\space}%
\providecommand \EOS [0]{\spacefactor3000\relax}%
\providecommand \BibitemShut  [1]{\csname bibitem#1\endcsname}%
\let\auto@bib@innerbib\@empty
\bibitem [{\citenamefont {Ekert}(1991)}]{PhysRevLett.67.661}%
  \BibitemOpen
  \bibfield  {author} {\bibinfo {author} {\bibfnamefont {A.~K.}\ \bibnamefont
  {Ekert}},\ }\href {\doibase 10.1103/PhysRevLett.67.661} {\bibfield  {journal}
  {\bibinfo  {journal} {Phys. Rev. Lett.}\ }\textbf {\bibinfo {volume} {67}},\
  \bibinfo {pages} {661} (\bibinfo {year} {1991})}\BibitemShut {NoStop}%
\bibitem [{\citenamefont {Bennett}\ \emph {et~al.}(1993)\citenamefont
  {Bennett}, \citenamefont {Brassard}, \citenamefont {Cr\'epeau}, \citenamefont
  {Jozsa}, \citenamefont {Peres},\ and\ \citenamefont
  {Wootters}}]{PhysRevLett.70.1895}%
  \BibitemOpen
  \bibfield  {author} {\bibinfo {author} {\bibfnamefont {C.~H.}\ \bibnamefont
  {Bennett}}, \bibinfo {author} {\bibfnamefont {G.}~\bibnamefont {Brassard}},
  \bibinfo {author} {\bibfnamefont {C.}~\bibnamefont {Cr\'epeau}}, \bibinfo
  {author} {\bibfnamefont {R.}~\bibnamefont {Jozsa}}, \bibinfo {author}
  {\bibfnamefont {A.}~\bibnamefont {Peres}}, \ and\ \bibinfo {author}
  {\bibfnamefont {W.~K.}\ \bibnamefont {Wootters}},\ }\href {\doibase
  10.1103/PhysRevLett.70.1895} {\bibfield  {journal} {\bibinfo  {journal}
  {Phys. Rev. Lett.}\ }\textbf {\bibinfo {volume} {70}},\ \bibinfo {pages}
  {1895} (\bibinfo {year} {1993})}\BibitemShut {NoStop}%
\bibitem [{\citenamefont {Bennett}\ and\ \citenamefont
  {Wiesner}(1992)}]{PhysRevLett.69.2881}%
  \BibitemOpen
  \bibfield  {author} {\bibinfo {author} {\bibfnamefont {C.~H.}\ \bibnamefont
  {Bennett}}\ and\ \bibinfo {author} {\bibfnamefont {S.~J.}\ \bibnamefont
  {Wiesner}},\ }\href {\doibase 10.1103/PhysRevLett.69.2881} {\bibfield
  {journal} {\bibinfo  {journal} {Phys. Rev. Lett.}\ }\textbf {\bibinfo
  {volume} {69}},\ \bibinfo {pages} {2881} (\bibinfo {year}
  {1992})}\BibitemShut {NoStop}%
\bibitem [{\citenamefont {Vedral}\ \emph {et~al.}(1997)\citenamefont {Vedral},
  \citenamefont {Plenio}, \citenamefont {Rippin},\ and\ \citenamefont
  {Knight}}]{PhysRevLett.78.2275}%
  \BibitemOpen
  \bibfield  {author} {\bibinfo {author} {\bibfnamefont {V.}~\bibnamefont
  {Vedral}}, \bibinfo {author} {\bibfnamefont {M.~B.}\ \bibnamefont {Plenio}},
  \bibinfo {author} {\bibfnamefont {M.~A.}\ \bibnamefont {Rippin}}, \ and\
  \bibinfo {author} {\bibfnamefont {P.~L.}\ \bibnamefont {Knight}},\ }\href
  {\doibase 10.1103/PhysRevLett.78.2275} {\bibfield  {journal} {\bibinfo
  {journal} {Phys. Rev. Lett.}\ }\textbf {\bibinfo {volume} {78}},\ \bibinfo
  {pages} {2275} (\bibinfo {year} {1997})}\BibitemShut {NoStop}%
\bibitem [{\citenamefont {Bru\ss}(2002)}]{10.1063/1.1494474}%
  \BibitemOpen
  \bibfield  {author} {\bibinfo {author} {\bibfnamefont {D.}~\bibnamefont
  {Bru\ss}},\ }\href {\doibase 10.1063/1.1494474} {\bibfield  {journal}
  {\bibinfo  {journal} {J. Math. Phys.}\ }\textbf {\bibinfo {volume} {43}},\
  \bibinfo {pages} {4237} (\bibinfo {year} {2002})}\BibitemShut {NoStop}%
\bibitem [{\citenamefont {Hill}\ and\ \citenamefont
  {Wootters}(1997)}]{PhysRevLett.78.5022}%
  \BibitemOpen
  \bibfield  {author} {\bibinfo {author} {\bibfnamefont {S.~A.}\ \bibnamefont
  {Hill}}\ and\ \bibinfo {author} {\bibfnamefont {W.~K.}\ \bibnamefont
  {Wootters}},\ }\href {\doibase 10.1103/PhysRevLett.78.5022} {\bibfield
  {journal} {\bibinfo  {journal} {Phys. Rev. Lett.}\ }\textbf {\bibinfo
  {volume} {78}},\ \bibinfo {pages} {5022} (\bibinfo {year}
  {1997})}\BibitemShut {NoStop}%
\bibitem [{\citenamefont {Eisert}\ and\ \citenamefont
  {Plenio}(1999)}]{doi:10.1080/09500349908231260}%
  \BibitemOpen
  \bibfield  {author} {\bibinfo {author} {\bibfnamefont {J.}~\bibnamefont
  {Eisert}}\ and\ \bibinfo {author} {\bibfnamefont {M.~B.}\ \bibnamefont
  {Plenio}},\ }\href {\doibase 10.1080/09500349908231260} {\bibfield  {journal}
  {\bibinfo  {journal} {J. Mod. Opt.}\ }\textbf {\bibinfo {volume} {46}},\
  \bibinfo {pages} {145} (\bibinfo {year} {1999})}\BibitemShut {NoStop}%
\bibitem [{\citenamefont {Vidal}\ and\ \citenamefont
  {Werner}(2002)}]{PhysRevA.65.032314}%
  \BibitemOpen
  \bibfield  {author} {\bibinfo {author} {\bibfnamefont {G.}~\bibnamefont
  {Vidal}}\ and\ \bibinfo {author} {\bibfnamefont {R.~F.}\ \bibnamefont
  {Werner}},\ }\href {\doibase 10.1103/PhysRevA.65.032314} {\bibfield
  {journal} {\bibinfo  {journal} {Phys. Rev. A}\ }\textbf {\bibinfo {volume}
  {65}},\ \bibinfo {pages} {032314} (\bibinfo {year} {2002})}\BibitemShut
  {NoStop}%
\bibitem [{\citenamefont {Singh}\ \emph {et~al.}(2020)\citenamefont {Singh},
  \citenamefont {Ahamed}, \citenamefont {Home},\ and\ \citenamefont
  {Sinha}}]{Singh:20}%
  \BibitemOpen
  \bibfield  {author} {\bibinfo {author} {\bibfnamefont {A.}~\bibnamefont
  {Singh}}, \bibinfo {author} {\bibfnamefont {I.}~\bibnamefont {Ahamed}},
  \bibinfo {author} {\bibfnamefont {D.}~\bibnamefont {Home}}, \ and\ \bibinfo
  {author} {\bibfnamefont {U.}~\bibnamefont {Sinha}},\ }\href {\doibase
  10.1364/JOSAB.37.000157} {\bibfield  {journal} {\bibinfo  {journal} {J. Opt.
  Soc. Am. B}\ }\textbf {\bibinfo {volume} {37}},\ \bibinfo {pages} {157}
  (\bibinfo {year} {2020})}\BibitemShut {NoStop}%
\bibitem [{\citenamefont {Hillery}\ \emph {et~al.}(1999)\citenamefont
  {Hillery}, \citenamefont {Bu\ifmmode~\check{z}\else \v{z}\fi{}ek},\ and\
  \citenamefont {Berthiaume}}]{PhysRevA.59.1829}%
  \BibitemOpen
  \bibfield  {author} {\bibinfo {author} {\bibfnamefont {M.}~\bibnamefont
  {Hillery}}, \bibinfo {author} {\bibfnamefont {V.}~\bibnamefont
  {Bu\ifmmode~\check{z}\else \v{z}\fi{}ek}}, \ and\ \bibinfo {author}
  {\bibfnamefont {A.}~\bibnamefont {Berthiaume}},\ }\href {\doibase
  10.1103/PhysRevA.59.1829} {\bibfield  {journal} {\bibinfo  {journal} {Phys.
  Rev. A}\ }\textbf {\bibinfo {volume} {59}},\ \bibinfo {pages} {1829}
  (\bibinfo {year} {1999})}\BibitemShut {NoStop}%
\bibitem [{\citenamefont {D\"ur}\ \emph {et~al.}(2000)\citenamefont {D\"ur},
  \citenamefont {Vidal},\ and\ \citenamefont {Cirac}}]{PhysRevA.62.062314}%
  \BibitemOpen
  \bibfield  {author} {\bibinfo {author} {\bibfnamefont {W.}~\bibnamefont
  {D\"ur}}, \bibinfo {author} {\bibfnamefont {G.}~\bibnamefont {Vidal}}, \ and\
  \bibinfo {author} {\bibfnamefont {J.~I.}\ \bibnamefont {Cirac}},\ }\href
  {\doibase 10.1103/PhysRevA.62.062314} {\bibfield  {journal} {\bibinfo
  {journal} {Phys. Rev. A}\ }\textbf {\bibinfo {volume} {62}},\ \bibinfo
  {pages} {062314} (\bibinfo {year} {2000})}\BibitemShut {NoStop}%
\bibitem [{\citenamefont {Barnum}\ and\ \citenamefont
  {Linden}(2001)}]{HBarnum_2001}%
  \BibitemOpen
  \bibfield  {author} {\bibinfo {author} {\bibfnamefont {H.}~\bibnamefont
  {Barnum}}\ and\ \bibinfo {author} {\bibfnamefont {N.}~\bibnamefont
  {Linden}},\ }\href {\doibase 10.1088/0305-4470/34/35/305} {\bibfield
  {journal} {\bibinfo  {journal} {J. Phys. A-Math. Gen.}\ }\textbf {\bibinfo
  {volume} {34}},\ \bibinfo {pages} {6787} (\bibinfo {year}
  {2001})}\BibitemShut {NoStop}%
\bibitem [{\citenamefont {Eisert}\ and\ \citenamefont
  {Briegel}(2001)}]{PhysRevA.64.022306}%
  \BibitemOpen
  \bibfield  {author} {\bibinfo {author} {\bibfnamefont {J.}~\bibnamefont
  {Eisert}}\ and\ \bibinfo {author} {\bibfnamefont {H.~J.}\ \bibnamefont
  {Briegel}},\ }\href {\doibase 10.1103/PhysRevA.64.022306} {\bibfield
  {journal} {\bibinfo  {journal} {Phys. Rev. A}\ }\textbf {\bibinfo {volume}
  {64}},\ \bibinfo {pages} {022306} (\bibinfo {year} {2001})}\BibitemShut
  {NoStop}%
\bibitem [{\citenamefont {Meyer}\ and\ \citenamefont
  {Wallach}(2002)}]{10.1063/1.1497700}%
  \BibitemOpen
  \bibfield  {author} {\bibinfo {author} {\bibfnamefont {D.~A.}\ \bibnamefont
  {Meyer}}\ and\ \bibinfo {author} {\bibfnamefont {N.~R.}\ \bibnamefont
  {Wallach}},\ }\href {\doibase 10.1063/1.1497700} {\bibfield  {journal}
  {\bibinfo  {journal} {J. Math. Phys.}\ }\textbf {\bibinfo {volume} {43}},\
  \bibinfo {pages} {4273} (\bibinfo {year} {2002})}\BibitemShut {NoStop}%
\bibitem [{\citenamefont {Carvalho}\ \emph {et~al.}(2004)\citenamefont
  {Carvalho}, \citenamefont {Mintert},\ and\ \citenamefont
  {Buchleitner}}]{PhysRevLett.93.230501}%
  \BibitemOpen
  \bibfield  {author} {\bibinfo {author} {\bibfnamefont {A.~R.~R.}\
  \bibnamefont {Carvalho}}, \bibinfo {author} {\bibfnamefont {F.}~\bibnamefont
  {Mintert}}, \ and\ \bibinfo {author} {\bibfnamefont {A.}~\bibnamefont
  {Buchleitner}},\ }\href {\doibase 10.1103/PhysRevLett.93.230501} {\bibfield
  {journal} {\bibinfo  {journal} {Phys. Rev. Lett.}\ }\textbf {\bibinfo
  {volume} {93}},\ \bibinfo {pages} {230501} (\bibinfo {year}
  {2004})}\BibitemShut {NoStop}%
\bibitem [{\citenamefont {Coffman}\ \emph {et~al.}(2000)\citenamefont
  {Coffman}, \citenamefont {Kundu},\ and\ \citenamefont
  {Wootters}}]{PhysRevA.61.052306}%
  \BibitemOpen
  \bibfield  {author} {\bibinfo {author} {\bibfnamefont {V.}~\bibnamefont
  {Coffman}}, \bibinfo {author} {\bibfnamefont {J.}~\bibnamefont {Kundu}}, \
  and\ \bibinfo {author} {\bibfnamefont {W.~K.}\ \bibnamefont {Wootters}},\
  }\href {\doibase 10.1103/PhysRevA.61.052306} {\bibfield  {journal} {\bibinfo
  {journal} {Phys. Rev. A}\ }\textbf {\bibinfo {volume} {61}},\ \bibinfo
  {pages} {052306} (\bibinfo {year} {2000})}\BibitemShut {NoStop}%
\bibitem [{\citenamefont {Jungnitsch}\ \emph {et~al.}(2011)\citenamefont
  {Jungnitsch}, \citenamefont {Moroder},\ and\ \citenamefont
  {G\"uhne}}]{PhysRevLett.106.190502}%
  \BibitemOpen
  \bibfield  {author} {\bibinfo {author} {\bibfnamefont {B.}~\bibnamefont
  {Jungnitsch}}, \bibinfo {author} {\bibfnamefont {T.}~\bibnamefont {Moroder}},
  \ and\ \bibinfo {author} {\bibfnamefont {O.}~\bibnamefont {G\"uhne}},\ }\href
  {\doibase 10.1103/PhysRevLett.106.190502} {\bibfield  {journal} {\bibinfo
  {journal} {Phys. Rev. Lett.}\ }\textbf {\bibinfo {volume} {106}},\ \bibinfo
  {pages} {190502} (\bibinfo {year} {2011})}\BibitemShut {NoStop}%
\bibitem [{\citenamefont {Ma}\ \emph {et~al.}(2011)\citenamefont {Ma},
  \citenamefont {Chen}, \citenamefont {Chen}, \citenamefont {Spengler},
  \citenamefont {Gabriel},\ and\ \citenamefont {Huber}}]{PhysRevA.83.062325}%
  \BibitemOpen
  \bibfield  {author} {\bibinfo {author} {\bibfnamefont {Z.-H.}\ \bibnamefont
  {Ma}}, \bibinfo {author} {\bibfnamefont {Z.-H.}\ \bibnamefont {Chen}},
  \bibinfo {author} {\bibfnamefont {J.-L.}\ \bibnamefont {Chen}}, \bibinfo
  {author} {\bibfnamefont {C.}~\bibnamefont {Spengler}}, \bibinfo {author}
  {\bibfnamefont {A.}~\bibnamefont {Gabriel}}, \ and\ \bibinfo {author}
  {\bibfnamefont {M.}~\bibnamefont {Huber}},\ }\href {\doibase
  10.1103/PhysRevA.83.062325} {\bibfield  {journal} {\bibinfo  {journal} {Phys.
  Rev. A}\ }\textbf {\bibinfo {volume} {83}},\ \bibinfo {pages} {062325}
  (\bibinfo {year} {2011})}\BibitemShut {NoStop}%
\bibitem [{\citenamefont {Sen(De)}\ and\ \citenamefont
  {Sen}(2010)}]{PhysRevA.81.012308}%
  \BibitemOpen
  \bibfield  {author} {\bibinfo {author} {\bibfnamefont {A.}~\bibnamefont
  {Sen(De)}}\ and\ \bibinfo {author} {\bibfnamefont {U.}~\bibnamefont {Sen}},\
  }\href {\doibase 10.1103/PhysRevA.81.012308} {\bibfield  {journal} {\bibinfo
  {journal} {Phys. Rev. A}\ }\textbf {\bibinfo {volume} {81}},\ \bibinfo
  {pages} {012308} (\bibinfo {year} {2010})}\BibitemShut {NoStop}%
\bibitem [{\citenamefont {Dong}\ \emph {et~al.}(2022)\citenamefont {Dong},
  \citenamefont {Wei}, \citenamefont {Song}, \citenamefont {Wang},\ and\
  \citenamefont {Ye}}]{PhysRevA.106.042415}%
  \BibitemOpen
  \bibfield  {author} {\bibinfo {author} {\bibfnamefont {D.-D.}\ \bibnamefont
  {Dong}}, \bibinfo {author} {\bibfnamefont {G.-B.}\ \bibnamefont {Wei}},
  \bibinfo {author} {\bibfnamefont {X.-K.}\ \bibnamefont {Song}}, \bibinfo
  {author} {\bibfnamefont {D.}~\bibnamefont {Wang}}, \ and\ \bibinfo {author}
  {\bibfnamefont {L.}~\bibnamefont {Ye}},\ }\href {\doibase
  10.1103/PhysRevA.106.042415} {\bibfield  {journal} {\bibinfo  {journal}
  {Phys. Rev. A}\ }\textbf {\bibinfo {volume} {106}},\ \bibinfo {pages}
  {042415} (\bibinfo {year} {2022})}\BibitemShut {NoStop}%
\bibitem [{\citenamefont {Dong}\ \emph {et~al.}(2023)\citenamefont {Dong},
  \citenamefont {Song}, \citenamefont {Fan}, \citenamefont {Ye},\ and\
  \citenamefont {Wang}}]{PhysRevA.107.052403}%
  \BibitemOpen
  \bibfield  {author} {\bibinfo {author} {\bibfnamefont {D.-D.}\ \bibnamefont
  {Dong}}, \bibinfo {author} {\bibfnamefont {X.-K.}\ \bibnamefont {Song}},
  \bibinfo {author} {\bibfnamefont {X.-G.}\ \bibnamefont {Fan}}, \bibinfo
  {author} {\bibfnamefont {L.}~\bibnamefont {Ye}}, \ and\ \bibinfo {author}
  {\bibfnamefont {D.}~\bibnamefont {Wang}},\ }\href {\doibase
  10.1103/PhysRevA.107.052403} {\bibfield  {journal} {\bibinfo  {journal}
  {Phys. Rev. A}\ }\textbf {\bibinfo {volume} {107}},\ \bibinfo {pages}
  {052403} (\bibinfo {year} {2023})}\BibitemShut {NoStop}%
\bibitem [{\citenamefont {Xie}\ and\ \citenamefont
  {Eberly}(2021)}]{PhysRevLett.127.040403}%
  \BibitemOpen
  \bibfield  {author} {\bibinfo {author} {\bibfnamefont {S.}~\bibnamefont
  {Xie}}\ and\ \bibinfo {author} {\bibfnamefont {J.~H.}\ \bibnamefont
  {Eberly}},\ }\href {\doibase 10.1103/PhysRevLett.127.040403} {\bibfield
  {journal} {\bibinfo  {journal} {Phys. Rev. Lett.}\ }\textbf {\bibinfo
  {volume} {127}},\ \bibinfo {pages} {040403} (\bibinfo {year}
  {2021})}\BibitemShut {NoStop}%
\bibitem [{\citenamefont {Li}\ and\ \citenamefont
  {Shang}(2022)}]{PhysRevResearch.4.023059}%
  \BibitemOpen
  \bibfield  {author} {\bibinfo {author} {\bibfnamefont {Y.}~\bibnamefont
  {Li}}\ and\ \bibinfo {author} {\bibfnamefont {J.}~\bibnamefont {Shang}},\
  }\href {\doibase 10.1103/PhysRevResearch.4.023059} {\bibfield  {journal}
  {\bibinfo  {journal} {Phys. Rev. Res.}\ }\textbf {\bibinfo {volume} {4}},\
  \bibinfo {pages} {023059} (\bibinfo {year} {2022})}\BibitemShut {NoStop}%
\bibitem [{\citenamefont {Ge}\ \emph {et~al.}(2023)\citenamefont {Ge},
  \citenamefont {Liu},\ and\ \citenamefont {Cheng}}]{PhysRevA.107.032405}%
  \BibitemOpen
  \bibfield  {author} {\bibinfo {author} {\bibfnamefont {X.}~\bibnamefont
  {Ge}}, \bibinfo {author} {\bibfnamefont {L.}~\bibnamefont {Liu}}, \ and\
  \bibinfo {author} {\bibfnamefont {S.}~\bibnamefont {Cheng}},\ }\href
  {\doibase 10.1103/PhysRevA.107.032405} {\bibfield  {journal} {\bibinfo
  {journal} {Phys. Rev. A}\ }\textbf {\bibinfo {volume} {107}},\ \bibinfo
  {pages} {032405} (\bibinfo {year} {2023})}\BibitemShut {NoStop}%
\bibitem [{\citenamefont {Ac\'{\i}n}\ \emph {et~al.}(2000)\citenamefont
  {Ac\'{\i}n}, \citenamefont {Andrianov}, \citenamefont {Costa}, \citenamefont
  {Jan\'e}, \citenamefont {Latorre},\ and\ \citenamefont
  {Tarrach}}]{PhysRevLett.85.1560}%
  \BibitemOpen
  \bibfield  {author} {\bibinfo {author} {\bibfnamefont {A.}~\bibnamefont
  {Ac\'{\i}n}}, \bibinfo {author} {\bibfnamefont {A.}~\bibnamefont
  {Andrianov}}, \bibinfo {author} {\bibfnamefont {L.}~\bibnamefont {Costa}},
  \bibinfo {author} {\bibfnamefont {E.}~\bibnamefont {Jan\'e}}, \bibinfo
  {author} {\bibfnamefont {J.~I.}\ \bibnamefont {Latorre}}, \ and\ \bibinfo
  {author} {\bibfnamefont {R.}~\bibnamefont {Tarrach}},\ }\href {\doibase
  10.1103/PhysRevLett.85.1560} {\bibfield  {journal} {\bibinfo  {journal}
  {Phys. Rev. Lett.}\ }\textbf {\bibinfo {volume} {85}},\ \bibinfo {pages}
  {1560} (\bibinfo {year} {2000})}\BibitemShut {NoStop}%
\bibitem [{\citenamefont {Mintert}\ \emph {et~al.}(2005)\citenamefont
  {Mintert}, \citenamefont {Ku\ifmmode~\acute{s}\else \'{s}\fi{}},\ and\
  \citenamefont {Buchleitner}}]{PhysRevLett.95.260502}%
  \BibitemOpen
  \bibfield  {author} {\bibinfo {author} {\bibfnamefont {F.}~\bibnamefont
  {Mintert}}, \bibinfo {author} {\bibfnamefont {M.}~\bibnamefont
  {Ku\ifmmode~\acute{s}\else \'{s}\fi{}}}, \ and\ \bibinfo {author}
  {\bibfnamefont {A.}~\bibnamefont {Buchleitner}},\ }\href {\doibase
  10.1103/PhysRevLett.95.260502} {\bibfield  {journal} {\bibinfo  {journal}
  {Phys. Rev. Lett.}\ }\textbf {\bibinfo {volume} {95}},\ \bibinfo {pages}
  {260502} (\bibinfo {year} {2005})}\BibitemShut {NoStop}%
\bibitem [{\citenamefont {Joo}\ \emph {et~al.}(2003)\citenamefont {Joo},
  \citenamefont {Park}, \citenamefont {Oh},\ and\ \citenamefont
  {Kim}}]{Joo_2003}%
  \BibitemOpen
  \bibfield  {author} {\bibinfo {author} {\bibfnamefont {J.}~\bibnamefont
  {Joo}}, \bibinfo {author} {\bibfnamefont {Y.-J.}\ \bibnamefont {Park}},
  \bibinfo {author} {\bibfnamefont {S.}~\bibnamefont {Oh}}, \ and\ \bibinfo
  {author} {\bibfnamefont {J.}~\bibnamefont {Kim}},\ }\href {\doibase
  10.1088/1367-2630/5/1/136} {\bibfield  {journal} {\bibinfo  {journal} {New J.
  Phys.}\ }\textbf {\bibinfo {volume} {5}},\ \bibinfo {pages} {136} (\bibinfo
  {year} {2003})}\BibitemShut {NoStop}%
\bibitem [{\citenamefont {Pope}\ and\ \citenamefont
  {Milburn}(2003)}]{PhysRevA.67.052107}%
  \BibitemOpen
  \bibfield  {author} {\bibinfo {author} {\bibfnamefont {D.~T.}\ \bibnamefont
  {Pope}}\ and\ \bibinfo {author} {\bibfnamefont {G.~J.}\ \bibnamefont
  {Milburn}},\ }\href {\doibase 10.1103/PhysRevA.67.052107} {\bibfield
  {journal} {\bibinfo  {journal} {Phys. Rev. A}\ }\textbf {\bibinfo {volume}
  {67}},\ \bibinfo {pages} {052107} (\bibinfo {year} {2003})}\BibitemShut
  {NoStop}%
\bibitem [{\citenamefont {Osterloh}\ \emph {et~al.}(2008)\citenamefont
  {Osterloh}, \citenamefont {Siewert},\ and\ \citenamefont
  {Uhlmann}}]{PhysRevA.77.032310}%
  \BibitemOpen
  \bibfield  {author} {\bibinfo {author} {\bibfnamefont {A.}~\bibnamefont
  {Osterloh}}, \bibinfo {author} {\bibfnamefont {J.}~\bibnamefont {Siewert}}, \
  and\ \bibinfo {author} {\bibfnamefont {A.}~\bibnamefont {Uhlmann}},\ }\href
  {\doibase 10.1103/PhysRevA.77.032310} {\bibfield  {journal} {\bibinfo
  {journal} {Phys. Rev. A}\ }\textbf {\bibinfo {volume} {77}},\ \bibinfo
  {pages} {032310} (\bibinfo {year} {2008})}\BibitemShut {NoStop}%
\bibitem [{\citenamefont {Wootters}(1998)}]{PhysRevLett.80.2245}%
  \BibitemOpen
  \bibfield  {author} {\bibinfo {author} {\bibfnamefont {W.~K.}\ \bibnamefont
  {Wootters}},\ }\href {\doibase 10.1103/PhysRevLett.80.2245} {\bibfield
  {journal} {\bibinfo  {journal} {Phys. Rev. Lett.}\ }\textbf {\bibinfo
  {volume} {80}},\ \bibinfo {pages} {2245} (\bibinfo {year}
  {1998})}\BibitemShut {NoStop}%
\bibitem [{\citenamefont {Lohmayer}\ \emph {et~al.}(2006)\citenamefont
  {Lohmayer}, \citenamefont {Osterloh}, \citenamefont {Siewert},\ and\
  \citenamefont {Uhlmann}}]{PhysRevLett.97.260502}%
  \BibitemOpen
  \bibfield  {author} {\bibinfo {author} {\bibfnamefont {R.}~\bibnamefont
  {Lohmayer}}, \bibinfo {author} {\bibfnamefont {A.}~\bibnamefont {Osterloh}},
  \bibinfo {author} {\bibfnamefont {J.}~\bibnamefont {Siewert}}, \ and\
  \bibinfo {author} {\bibfnamefont {A.}~\bibnamefont {Uhlmann}},\ }\href
  {\doibase 10.1103/PhysRevLett.97.260502} {\bibfield  {journal} {\bibinfo
  {journal} {Phys. Rev. Lett.}\ }\textbf {\bibinfo {volume} {97}},\ \bibinfo
  {pages} {260502} (\bibinfo {year} {2006})}\BibitemShut {NoStop}%
\bibitem [{\citenamefont {Jung}\ \emph {et~al.}(2009)\citenamefont {Jung},
  \citenamefont {Hwang}, \citenamefont {Park},\ and\ \citenamefont
  {Son}}]{PhysRevA.79.024306}%
  \BibitemOpen
  \bibfield  {author} {\bibinfo {author} {\bibfnamefont {E.}~\bibnamefont
  {Jung}}, \bibinfo {author} {\bibfnamefont {M.-R.}\ \bibnamefont {Hwang}},
  \bibinfo {author} {\bibfnamefont {D.}~\bibnamefont {Park}}, \ and\ \bibinfo
  {author} {\bibfnamefont {J.-W.}\ \bibnamefont {Son}},\ }\href {\doibase
  10.1103/PhysRevA.79.024306} {\bibfield  {journal} {\bibinfo  {journal} {Phys.
  Rev. A}\ }\textbf {\bibinfo {volume} {79}},\ \bibinfo {pages} {024306}
  (\bibinfo {year} {2009})}\BibitemShut {NoStop}%
\end{thebibliography}

%

\end{document}